\documentclass[twocolumn]{aastex62}

\hypersetup{linkcolor=red,citecolor=blue,filecolor=cyan,urlcolor=cyan}
\usepackage{amsmath}
\usepackage{natbib,twoopt}

\newcommand{\dist}{720.35} 
\newcommand{\edist}{7.84}
\newcommand{\parallax}{1.388} 
\newcommand{\eparallax}{0.015}
\newcommand{\Mcep}{4.288} 
\newcommand{\eMcep}{0.133}
\newcommand{\Mcomp}{4.040} 
\newcommand{\eMcomp}{0.048}
\newcommand{\diamLD}{0.524} 
\newcommand{\ediamLD}{0.039}
\newcommand{\Mratio}{0.942} 
\newcommand{\eMratio}{0.020}
\newcommand{\Rcep}{40.6} 
\newcommand{\eRcep}{3.0}

\definecolor{red}{rgb}{1.,0.0,0.}

\shorttitle{A geometrical 1\,\% distance to a short-period binary Cepheid}
\shortauthors{Gallenne et al.}

\begin{document}

\title{\large \textbf{A geometrical 1\,\% distance to the short-period binary Cepheid V1334~Cygni}}

\correspondingauthor{Alexandre Gallenne}
\email{agallenn@eso.org}

\author[0000-0001-7853-4094]{A.~Gallenne}
\affil{European Southern Observatory, Alonso de C\'ordova 3107, Casilla 19001, Santiago, Chile}
\author[0000-0003-0626-1749]{P.~Kervella}
\affiliation{LESIA, Observatoire de Paris, Universit\'e PSL, CNRS, Sorbonne Universit\'e,\\
	Univ. Paris Diderot, Sorbonne Paris Cit\'e, 5 place Jules Janssen, 92195 Meudon, France}
\author[0000-0002-4374-075X]{N.~R.~Evans}
\affiliation{Smithsonian Astrophysical Observatory, MS 4, 60 Garden Street, Cambridge, MA 02138, USA}
\author[0000-0001-7617-5665]{C.R~Proffitt}
\affiliation{Space Telescope Science Institute, 3700 San Martin Drive, Baltimore, MD 21218, USA}
\author[0000-0002-3380-3307]{J.~D.~Monnier}
\affiliation{Astronomy Department, University of Michigan, 941 Dennison Bldg, Ann Arbor, MI 48109-1090, USA}
\author[0000-0003-2125-0183]{A.~M\'erand}
\affiliation{European Southern Observatory, Karl-Schwarzschild-Str. 2, 85748 Garching, Germany}
\author[0000-0002-5704-5221]{E.~Nelan}
\affiliation{Space Telescope Science Institute, 3700 San Martin Drive, Baltimore, MD 21218, USA}
\author[0000-0001-9065-6633]{E.~Winston}
\affiliation{Smithsonian Astrophysical Observatory, MS 4, 60 Garden Street, Cambridge, MA 02138, USA}
\author{G.~Pietrzy\'nski}
\affiliation{Centrum Astronomiczne im. Miko\l{}aja Kopernika, PAN, Bartycka 18, 00-716 Warsaw, Poland}
\author[0000-0001-5415-9189]{G.~Schaefer}
\affiliation{The CHARA Array of Georgia State University, Mount Wilson CA 91023, USA}
\author[0000-0003-1405-9954]{W.~Gieren}
\affiliation{Universidad de Concepci\'on, Departamento de Astronom\'ia, Casilla 160-C, Concepci\'on, Chile}
\affiliation{Millennium Institute of Astrophysics (MAS), Chile}
\author[0000-0001-8089-4419]{R.~I.~Anderson}
\affiliation{European Southern Observatory, Karl-Schwarzschild-Str. 2, 85748 Garching, Germany}
\author[0000-0002-4829-4955]{S.~Borgniet}
\affiliation{LESIA (UMR 8109), Observatoire de Paris, PSL, CNRS, UPMC, Univ. Paris-Diderot\\
	5 place Jules Janssen, 92195 Meudon, France}
\author[0000-0001-6017-8773]{S.~Kraus}
\affiliation{University of Exeter, School of Physics and Astronomy, Stocker Road, Exeter, EX4 4QL}
\author[0000-0002-9288-3482]{R.~M.~Roettenbacher}
\affiliation{Astronomy Department, University of Michigan, 941 Dennison Bldg, Ann Arbor, MI 48109-1090, USA}
\affiliation{Department of Astronomy, Stockholm University, SE-106 91 Stockholm, Sweden}
\author[0000-0002-8376-8941]{F.~Baron}
\affiliation{Astronomy Department, University of Michigan, 941 Dennison Bldg, Ann Arbor, MI 48109-1090, USA}
\affiliation{Department of Physics and Astronomy, Georgia State University, Atlanta, GA 30303, USA}
\author[0000-0003-3861-8124]{B.~Pilecki}
\affiliation{Centrum Astronomiczne im. Miko\l{}aja Kopernika, PAN, Bartycka 18, 00-716 Warsaw, Poland}
\author[0000-0002-1560-8620]{M.~Taormina}
\affiliation{Centrum Astronomiczne im. Miko\l{}aja Kopernika, PAN, Bartycka 18, 00-716 Warsaw, Poland}
\author[0000-0002-7355-9775]{D.~Graczyk}
\affiliation{Centrum Astronomiczne im. Miko\l{}aja Kopernika, PAN, Rabia\'nska 8, 87–100 Toru\'n, Poland}
\author[0000-0003-1578-6993]{N.~Mowlavi}
\affiliation{D\'epartement d'Astronomie, Universit\'e de Gen\`eve, 51 Ch. des Maillettes, 1290 Sauverny, Switzerland}
\author[0000-0002-0182-8040]{L.~Eyer}
\affiliation{D\'epartement d'Astronomie, Universit\'e de Gen\`eve, 51 Ch. des Maillettes, 1290 Sauverny, Switzerland}



\begin{abstract}
Cepheid stars play a considerable role as extragalactic distances indicators, thanks to the simple empirical relation between their pulsation period and their luminosity. They overlap with that of secondary distance indicators, such as Type Ia supernovae, whose distance scale is tied to Cepheid luminosities. However, the Period--Luminosity (P-L) relation still lacks a calibration to better than 5\,\%. Using an original combination of interferometric astrometry with optical and ultraviolet spectroscopy, we measured the geometrical distance $d = \dist \pm \edist$\,pc of the 3.33\,d period Cepheid V1334 Cyg with an unprecedented accuracy of $\pm 1$\,\%, providing the most accurate distance for a Cepheid. Placing this star in the P--L diagram provides an independent test of existing period-luminosity relations. We show that the secondary star has a significant impact on the integrated magnitude, particularly at visible wavelengths. Binarity in future high precision calibrations of the P--L relations is not negligible, at least in the short-period regime. Subtracting the companion flux leaves V1334 Cyg in marginal agreement with existing photometric-based P--L relations, indicating either an overall calibration bias or a significant intrinsic dispersion at a few percent level. Our work also enabled us to determine the dynamical masses of both components, $M_1 = \Mcep \pm \eMcep\,M_\odot$ (Cepheid) and $M_2 = \Mcomp \pm \eMcomp\,M_\odot$ (companion), providing the most accurate masses for a Galactic binary Cepheid system.
\end{abstract}



\section{Introduction}
\label{section__introduction}

	Cepheids play a particularly important role as extragalactic distance indicators thanks to the empirical relation between their pulsation period and their luminosity: the Period-Luminosity (P--L) relation or Leavitt law \citep{Leavitt_1912_03_0}. The ability to predict their intrinsic luminosity from their variation period makes them precious standard candles, as its combination with their measured apparent magnitude gives the distance modulus in a simple, empirical manner. Cepheids are radially pulsating, young and bright supergiant stars that are easily identified in distant galaxies ($\lesssim 25$\,Mpc). Their distance scale overlaps with secondary distance indicators, such as supernovae Ia (SNIa), whose distance scale is securely tied to Cepheid luminosities. Therefore, an accurate and precise calibration of the Cepheid luminosities is necessary to ensure a reliable calibration for the secondary distance indicators. However, despite considerable efforts over more than one century, the calibration of the zero point (ZP) of the P--L relation has not yet reached a 1\% accuracy level. A better control of the systematic uncertainties is needed to establish the P--L relation calibration at this level, and secondary effects on the photometric measurements such as a metallicity-dependence or interstellar reddening must be understood and taken into consideration.
	
	It is now clear that the intrinsic scatter of the relation is smaller at longer wavelengths \citep[see e.g.][]{Freedman_2010_09_0,Madore_2012_01_0} than it is for $V$ band ($\sim 0.5\,\mu$m) measurements. In optical bands, the line-of-sight extinction is stronger than in the infrared, and  flux contamination from any main-sequence companions is likely to be larger -- from an evolutionary time-scale point of view, most of the companions should be stars close to the main-sequence\citep[see e.g.][]{Bohm-Vitense_1985_09_1,Evans_1992_01_0,Szabados_2012_09_0}. As they are relatively massive stars, the binary fraction of Cepheids is known to be at least 50\,\% \citep{Szabados_2003__0}, and the presence of companions may bias the calibration of the P--L relation if their photometric contribution is not removed \citep{Gaia-Collaboration_2017_09_0,Szabados_2012_09_0}. Furthermore, this bias is likely not constant as a function of the pulsation period as the bright long-period Cepheids (hence massive) have a higher contrast with their companions (reducing the potential photometric bias) compared to the shortest periods (less massive), but they are also more likely to be members of multiple stars systems with more massive companions than are short period Cepheids \cite[see e.g.][increasing the bias]{Sana_2011_07_0}. In the near- and mid-IR,  Cepheids usually strongly overshine their companions, and such effects are reduced. On the other hand, the presence of circumstellar envelopes increases the measured flux beyond that of the Cepheid itself ($\lesssim 5$\,\% and $1-30$\,\% of IR excess in the near- and mid-IR, respectively), and so systematically affects the distance estimates \citep{Gallenne_2017_11_0,Gallenne_2013_10_0,Kervella_2006_03_0,Merand_2006_07_0}. The optimal band seems to be the $K$ band ($\sim 2.2\,\mu$m) where the contributions of both the companions and the envelopes are mitigated, but the Cepheid distance scale must nevertheless be tested at the percent level. Such test seems to be even more critical with the growing evidence of a $\sim 3.7\sigma$ discrepancy between the value of the Hubble constant $H_0$ predicted from Cosmic Microwave Background anisotropies \citep{Planck-Collaboration_2018_07_0} and the recent empirical calibration using Cepheids and SNIa \citep{Riess_2016_07_0}. Before invoking new physics in the standard cosmological model, new high accuracy data are necessary.
	
	
	For now, the calibration of the P--L relation is partially anchored to 17 Galactic Cepheids having trigonometric parallax measurements, obtained from the Hubble Space Telescope (HST) Fine Guidance Sensor \citep[][10 Cepheids]{Benedict_2007_04_0,Benedict_2002_09_0} and Wide Field Camera 3 spatial scanning \citep[][7 Cepheids]{Riess_2018_03_0}. Unfortunately, their average accuracy is limited to 8\,\%, preventing a reliable 1\,\% calibration of their luminosity. While the final Gaia data release expected in 2022 will provide hundreds of Cepheid parallaxes to 3\% or less, our knowledge of the interstellar extinction will still be an important limitation to derive highly accurate Cepheid luminosities. In addition, close orbiting companions could bias the Gaia astrometric measurements and therefore the resulting parallaxes if they are assumed single stars \citep[see e.g.][]{Anderson_2016_10_0}. Here we report the determination of the geometrical distance of the Galactic binary Cepheid V1334~Cyg with a very high accuracy of 1\,\%, providing the most accurate anchor point of the P--L relation for short period Cepheids. Our observations also enable us to confirm and measure the bias due to the companion on P--L calibrations.

	Binary Cepheids offer the opportunity to measure accurate distances and masses through the combination of astrometric (visual orbit) and spectroscopic (radial velocities) measurements. But this is a challenging task because of 1) the small angular size of the orbits that hampers astrometric measurements using single-dish telescopes, and 2) the high contrast between the Cepheid and its companion, that complicates the detection of the companion's spectral lines and the measurement of its radial velocity. Despite these difficulties, we report in the present work the detection of both the Cepheid and secondary component of V1334 Cyg using interferometric, as well as visible and ultraviolet spectroscopic observations. This system was the first target chosen for our interferometric observing campaign, first because the expected $H$-band contrast between the components is reachable by the current interferometric combiners, and also because the spectroscopic orbit is particularly well known \citep{Evans_1992_01_0,Evans_2000_06_0}. We present in Sect.~\ref{section__observations_and_data_reduction} all the data we collected for the V1334~Cyg system. In Sect.~\ref{section__model_fitting_and_model_fitting} we introduce the models we used to fit our observations and derive our orbital solutions. We then discuss and conclude our results in Sect.~\ref{section__discussion_and_conclusions}.

\section{Observations and data reduction}
\label{section__observations_and_data_reduction}

	\begin{figure*}[ht]
		\centering
		\resizebox{\hsize}{!}{\includegraphics{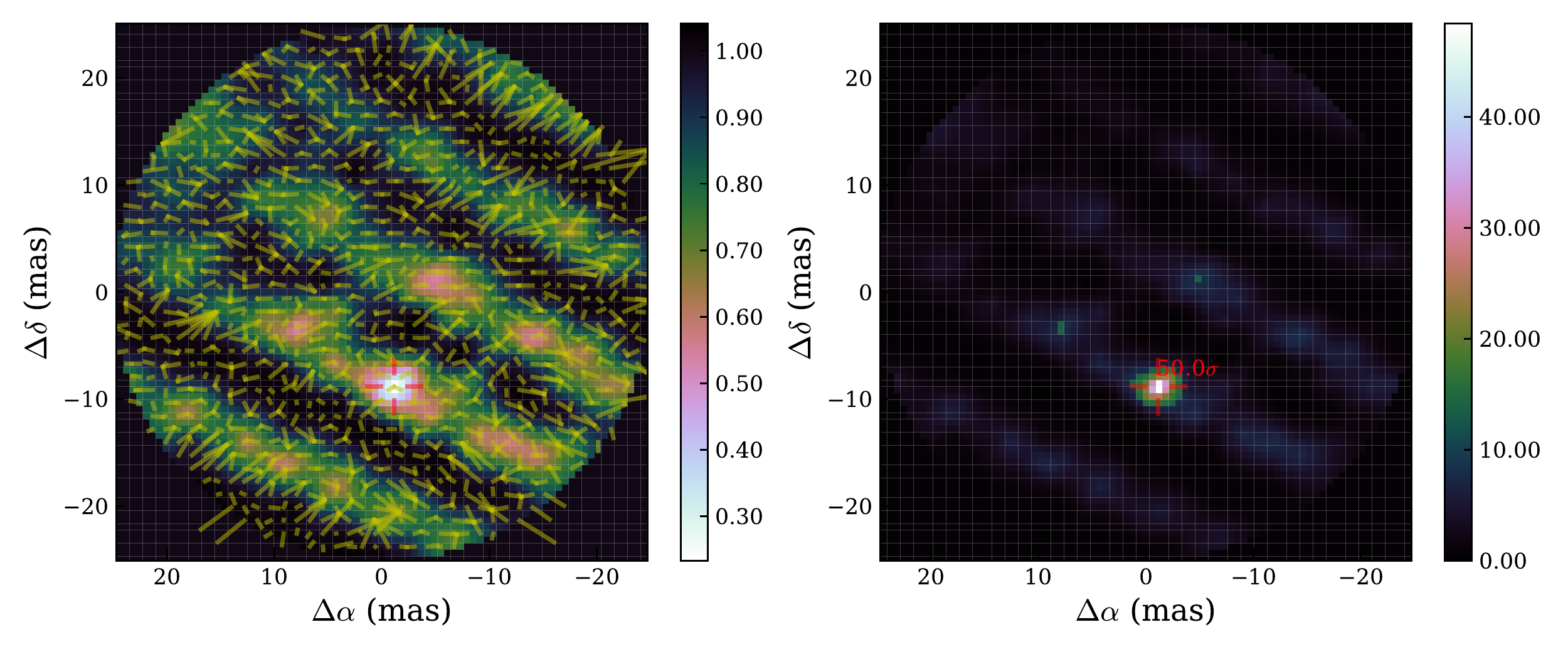}}
		\caption{$\chi^2_r$ map of the local minima (left) and detection level map (right) of V1334 Cyg for the July 2012 observations. The yellow lines represent the convergence from the starting points to the final fitted position, as explained in \citet{Gallenne_2015_07_0}. The maps were reinterpolated in a regular grid for clarity. The axis limits were chosen according to the location of the companion.}
		\label{figure_chi2map}
	\end{figure*}

\subsection{Interferometry}
	The relative astrometric positions of the companion of V1334~Cyg were measured repeatedly using interferometric observations performed with the Michigan InfraRed Combiner \citep{Monnier_2004_10_0}, installed at the Center for High Angular Resolution Astronomy array \citep{ten-Brummelaar_2005_07_0} located on Mount Wilson, California. The array consists of six 1\,m aperture telescopes with an Y-shaped configuration (two telescopes on each branch), oriented to the east, west and south, and so offering a good coverage of the $(u, v)$ plane. The baselines range from 34\,m to 331\,m, providing an angular resolution down to 0.5\,mas in $H$. The MIRC instrument combines the light coming from all six telescopes in the $H$ band, with three spectral resolutions ($R = 42, 150$ and 400). The recombination of six telescopes gives simultaneously 15 fringe visibilities and 20 closure phase measurements, that are our primary observables. Our observations were carried out from 2012 July to 2016 October using either four, five or six telescopes. We used the low spectral resolution mode, where a prism splits the light on the detector into 8 narrow spectral channels. We followed a standard observing procedure, i.e. we monitored the interferometric transfer function by observing a calibrator before and after our Cepheids. The calibrators were selected using the SearchCal software\footnote{Available at \url{http://www.jmmc.fr/searchcal}.} \citep{Bonneau_2006_09_0} provided by the Jean-Marie Mariotti Center (JMMC). They are listed in Table~\ref{table_cal}.

	\begin{table}[h]
		\centering
		\caption{$H$-band uniform disk sizes of calibrators.} 
		\begin{tabular}{cccc}
			\hline
			\hline
			Star  & $\theta_\mathrm{UD}$  & $V$  &  $H$  \\
					& 	(mas)							& (mag) & (mag)  \\
			\hline
			HD192985	&	$0.450\pm0.032$	   & 5.87  & 4.87  \\ 
			HD199956	&	$0.603\pm0.043$	   & 6.65  & 4.46 \\  
			HD202850	&	$0.566\pm0.049$	   & 4.22  & 3.86 \\  
			HD204153	&	$0.440\pm0.031$	   & 5.59  & 4.86 \\  
			HD207978	&	$0.571\pm0.040$    & 5.52  & 4.44 \\  
			HD212487	&	$0.434\pm0.031$	   & 6.18  & 4.97  \\  
			HD214200	&	$0.704\pm0.050$    & 6.11  & 4.16  \\  
			HD218470	&	$0.477\pm0.034$    & 6.68  & 4.67  \\  
			\hline
		\end{tabular}
		\label{table_cal}
	\end{table}

	The data were reduced with the standard MIRC pipeline \citep{Monnier_2007_07_0}. The main procedure is to compute squared visibilities and triple products for each baseline and spectral channel, and to correct for photon and readout noises.

	For each epoch, astrometric position were determined by fitting squared visibilities ($V^2$) and closure phases ($CP$) with our dedicated tool \texttt{CANDID} \citep{Gallenne_2015_07_0}, with errors estimated using the bootstrapping technique (with replacement and 1000 bootstrap sample). At each epoch, the fitted parameters are the relative astrometric position $(\Delta \alpha, \Delta \delta)$, the uniform disk (UD) angular diameter of the primary $\theta_\mathrm{UD}$, and the flux ratio between the components $f$. The equation for the complex visibility for a binary system with an unresolved companion is:
	\begin{eqnarray*}
		V(u,v) &=& \frac{V_\star(u,v) + f V_c(u,v)}{1 + f} \\
		V_\star(u,v) &=& \frac{2 J_1(\pi \theta_\mathrm{UD} B/\lambda)}{\pi \theta_\mathrm{UD} B/\lambda} \\
		V_c(u,v) &=& \exp [-\frac{2i\pi}{\lambda} (u\Delta \alpha + v\Delta \delta)] \\
		\phi_\mathrm{ijk} &=& \arg [V(u_i,v_i)V(u_j,v_j)V^*(u_k,v_k)],
	\end{eqnarray*}
	where $V_\star(u,v)$ and $V_c(u,v)$ are the complex visibilities of the Cepheid and companion, respectively, $J_1$ the first-order Bessel function, $(u, v)$ the spatial frequencies, $B$ the baseline between two telescopes $(i,j)$, $\lambda$ the observing wavelength, and $\phi_\mathrm{ijk}$ the closure phase for each closed baseline triangle $(i,j,k)$. More details about the companion search formalism with \texttt{CANDID} are available in \citet{Gallenne_2015_07_0,Gallenne_2013_04_0}. Briefly, the tool provides a 2D-grid of fit using a least-squares algorithm. For each starting point in the grid, a multi-parameter fit is performed, i.e. adjusting $\Delta \alpha, \Delta \delta, \theta_\mathrm{UD}$, and $f$. Each position of the grid leads to a local minimum. If the starting grid is fine enough (estimated a posteriori), multiple starting points lead to the same local minima, guaranteeing that all the local minima are explored. The deepest location identifies the best local minimum. Fig.~\ref{figure_chi2map} shows the map of the local minima of the reduced $\chi^2$ for our first observing epoch, in which the companion is detected at more than $50\sigma$. The measured astrometric positions are listed in Table~\ref{table_astrometry}.

	These observations also provides us the diameter of the Cepheid at various epochs of its pulsation phase. However, the angular diameter variation is at most 0.02\,mas for such a short period Cepheid. Visibility measurements with an accuracy better than 2\,\% would be required to detect such a variation, and this is well below the average MIRC accuracy for a 4.7\,mag star. We therefore determined the average UD diameter simply by calculating the mean and standard deviation of our measurements. The conversion from UD to limb-darkened (LD) angular diameter is done by using a linear-law parametrization with the LD coefficient $u_\lambda = 0.2423$, as explained in \citet{Gallenne_2013_04_0}. We measured $\theta_\mathrm{LD} = \diamLD \pm \ediamLD$\,mas. Combined with our measured distance, we estimate the linear radius of the Cepheid to be $R_1 = \Rcep \pm \eRcep\,\mathrm{R_\odot}$.

\begin{table}[ht]
	\centering
	\caption{Measured astrometric positions and flux ratio ($H$-band) of the companion of V1334 Cyg.} 
	\begin{tabular}{cccc}
		\hline
		\hline
		MJD  & $\Delta \alpha$	&	$\Delta \delta$	&	$f$    \\
		& 	(mas)					&  (mas)				&	(\%)    \\
		\hline
		56135.461  &   $-1.140\pm0.142$  &   $-8.832\pm0.043$  &   $3.39\pm0.60$  \\
		56201.211  &   $-0.099\pm0.029$  &   $-8.365\pm0.023$  &   $3.32\pm0.37$  \\
		56505.474  &   $4.450\pm0.116$  &   $-3.695\pm0.050$  &   $2.99\pm0.52$  \\
		56560.226  &   $4.935\pm0.008$  &   $-2.615\pm0.010$  &   $3.19\pm0.16$  \\
		56855.307  &   $5.818\pm0.023$  &   $3.786\pm0.024$  &   $2.89\pm0.26$  \\
		56860.413  &   $5.815\pm0.033$  &   $3.844\pm0.028$  &   $3.31\pm0.46$  \\
		56862.406  &   $5.744\pm0.028$  &   $3.937\pm0.025$  &   $3.31\pm0.15$  \\
		56931.307  &   $5.094\pm0.027$  &   $5.062\pm0.018$  &   $3.45\pm0.35$  \\
		57228.411  &   $-1.280\pm0.064$  &   $4.306\pm0.037$  &   $3.08\pm0.25$  \\
		57235.412  &   $-1.398\pm0.022$  &   $4.130\pm0.014$  &   $3.12\pm0.15$  \\
		57587.410  &   $-6.008\pm0.035$  &   $-4.987\pm0.042$  &   $4.26\pm0.58$  \\
		\hline
	\end{tabular}
	\label{table_astrometry}
\end{table}

\subsection{Visible spectroscopy}

	High quality optical spectra were collected using the fiber-fed high-resolution SOPHIE and HERMES spectrographs \citep{Bouchy_2009_10_0,Raskin_2011_02_0}. A total of 14 measurements have been obtained from 2013 to 2016 with SOPHIE, mounted on the 1.93\,m telescope of the Observatoire de Haute Provence (France). The instrument covers visible wavelengths with a spectral resolution $R \sim 75~000$ (high-resolution mode). We also collected 56 spectra from 2010 to 2017 with HERMES ($R \sim 85~000$), mounted on the Flemish 1.2\,m telescope of the Roque de los Muchachos Observatory (La Palma, Canary Islands). Exposure times of a few minutes allowed a signal-to-noise ratio per pixel at 550nm $>50$. Data were reduced using the dedicated HERMES and SOPHIE pipelines.

	Radial velocities (RVs) were estimated using the cross-correlation method. We created our own weighted binary mask by selecting unblended lines from high-resolution synthetic spectra ($R \sim 120~000$) covering the wavelength range 4500-6800\,\AA. The cross-correlation function is then fitted by a Gaussian whose minimum value gives an estimate of the radial velocity. Uncertainties include photon noise and internal drift. In addition, the standard deviation of the residual of the pulsation fit was also quadratically added. Measurements are listed in Table~\ref{table_rv1}. The zero point difference between the instruments has been estimated using the systemic velocity as follows. We first performed an orbital fit using only the SOPHIE data, then with only the HERMES ones. We measured a difference of 30\,m~s$^{-1}$, which we added to the SOPHIE measurements, taking Hermes data as the reference.

	\begin{table*}[ht]
	\centering
	\caption{Radial velocities of V1334~Cyg.} 
	\begin{tabular}{ccc|ccc}
		\hline
		\hline
		BJD-2400000.5  & $V_1$					& Instr. & BJD-2400000.5  & $V_1$	& Instr.   \\
				(day)		& 	(km~s$^{-1}$)	&           &    (day)                &   (km~s$^{-1}$)  &					    \\
		\hline
56496.018  &   $0.683\pm0.013$  &  SOPHIE  &  56493.168  &   $-2.082\pm0.017$  &  HERMES  \\
56497.065  &   $-7.105\pm0.014$  &  SOPHIE  &  56493.229  &   $-2.651\pm0.017$  &  HERMES  \\
56498.038  &   $-5.153\pm0.013$  &  SOPHIE  &  56493.917  &   $-7.496\pm0.017$  &  HERMES  \\
56500.021  &   $-4.386\pm0.014$  &  SOPHIE  &  56494.000  &   $-7.555\pm0.017$  &  HERMES  \\
56586.910  &   $-9.291\pm0.024$  &  SOPHIE  &  56494.201  &   $-7.175\pm0.017$  &  HERMES  \\
56587.880  &   $-8.900\pm0.013$  &  SOPHIE  &  56494.235  &   $-7.101\pm0.017$  &  HERMES  \\
56588.955  &   $-2.648\pm0.012$  &  SOPHIE  &  56494.908  &   $-3.740\pm0.016$  &  HERMES  \\
56829.089  &   $-10.946\pm0.011$  &  SOPHIE  &  56495.235  &   $-1.540\pm0.016$  &  HERMES  \\
56830.100  &   $-17.562\pm0.011$  &  SOPHIE  &  56495.958  &   $0.736\pm0.017$  &  HERMES  \\
56831.091  &   $-17.999\pm0.021$  &  SOPHIE  &  56496.055  &   $0.588\pm0.017$  &  HERMES  \\
57200.059  &   $-15.583\pm0.018$  &  SOPHIE  &  56496.237  &   $-0.136\pm0.017$  &  HERMES  \\
57201.084  &   $-14.880\pm0.012$  &  SOPHIE  &  56855.146  &   $-14.091\pm0.008$  &  HERMES  \\
57644.919  &   $10.503\pm0.011$  &  SOPHIE  &  56856.156  &   $-13.070\pm0.008$  &  HERMES  \\
57649.968  &   $5.864\pm0.018$  &  SOPHIE  &  56857.153  &   $-20.297\pm0.008$  &  HERMES  \\
55522.925  &   $7.657\pm0.017$  &  HERMES  &  56858.150  &   $-16.430\pm0.008$  &  HERMES  \\
56096.209  &   $10.746\pm0.017$  &  HERMES  &  56859.150  &   $-11.935\pm0.008$  &  HERMES  \\
56097.196  &   $2.917\pm0.017$  &  HERMES  &  56860.154  &   $-18.842\pm0.008$  &  HERMES  \\
56098.189  &   $5.372\pm0.016$  &  HERMES  &  56861.155  &   $-18.484\pm0.008$  &  HERMES  \\
56099.203  &   $10.961\pm0.016$  &  HERMES  &  56862.149  &   $-12.708\pm0.007$  &  HERMES  \\
56488.993  &   $0.782\pm0.016$  &  HERMES  &  56864.136  &   $-20.193\pm0.007$  &  HERMES  \\
56489.095  &   $1.116\pm0.016$  &  HERMES  &  56864.219  &   $-19.919\pm0.007$  &  HERMES  \\
56489.193  &   $1.258\pm0.016$  &  HERMES  &  56865.146  &   $-14.547\pm0.006$  &  HERMES  \\
56489.978  &   $-3.563\pm0.017$  &  HERMES  &  56865.223  &   $-14.083\pm0.007$  &  HERMES  \\
56490.062  &   $-4.360\pm0.017$  &  HERMES  &  56866.142  &   $-13.293\pm0.007$  &  HERMES  \\
56490.226  &   $-5.724\pm0.017$  &  HERMES  &  56866.220  &   $-13.844\pm0.008$  &  HERMES  \\
56490.951  &   $-7.091\pm0.017$  &  HERMES  &  57178.090  &   $-14.217\pm0.012$  &  HERMES  \\
56491.001  &   $-6.878\pm0.016$  &  HERMES  &  57179.098  &   $-9.793\pm0.011$  &  HERMES  \\
56491.039  &   $-6.721\pm0.016$  &  HERMES  &  57180.091  &   $-16.895\pm0.009$  &  HERMES  \\
56491.198  &   $-5.855\pm0.016$  &  HERMES  &  57181.155  &   $-15.587\pm0.007$  &  HERMES  \\
56491.934  &   $-1.301\pm0.016$  &  HERMES  &  57182.187  &   $-9.629\pm0.007$  &  HERMES  \\
56492.004  &   $-0.835\pm0.016$  &  HERMES  &  58026.968  &   $2.570\pm0.026$  &  HERMES  \\
56492.041  &   $-0.592\pm0.016$  &  HERMES  &  58035.037  &   $9.640\pm0.014$  &  HERMES  \\
56492.952  &   $-0.433\pm0.017$  &  HERMES  &  58070.845  &   $3.797\pm0.022$  &  HERMES  \\
56493.000  &   $-0.770\pm0.017$  &  HERMES  &  58071.917  &   $9.969\pm0.015$  &  HERMES  \\
56493.037  &   $-1.013\pm0.017$  &  HERMES  &  58073.878  &   $2.356\pm0.017$  &  HERMES  \\

		\hline
		\end{tabular}
		\label{table_rv1}
	\end{table*}

\subsection{Ultraviolet spectroscopy}

	Measuring the orbital velocity of the companion is also necessary to resolve the degeneracy between the distance and the masses. Due to its early spectral type, the companion is best detected at ultraviolet (UV) wavelengths, which is only possible with the HST because of the UV absorption by the Earth's atmosphere. Four observations were obtained with the Space Telescope Imaging Spectrograph (STIS) on board HST. We used the high resolution echelle mode E140H in the wavelength region 1163 to 1357\AA. They were taken over three annual HST cycles. Two were made very close in time to determine whether the companion is itself a short period binary. The exposure time for each was 1800 seconds. Data reductions was made in the same way as in \citet{Evans_2018_09_0}. Briefly, for each epoch, the 49 echelle orders were combined into a single one dimension spectrum using an appropriate blaze function. Then, interstellar lines, which are recognizably narrow compared with the stellar lines, were removed by interpolation.

	The absolute stability of the STIS E140H wavelength scale over time can be checked by comparing the wavelengths of interstellar features in different observations of the same star.  Using these observations of V1334~Cyg and similar observations of the hot companion to S~Mus, \citet{Proffitt_2017_09_0} found the dispersion in the zero-point of E140H velocity measurements to be about $0.24\,\mathrm{km~s^{-1}}$.

	Radial velocities of the companion were determined by cross-correlation. No synthetic template spectrum was used, instead we estimated a velocity difference with respect to the first observation by cross-correlating each of the spectra against it. This was typically done in 11 segments of approximately 10\,\AA, which were inspected to optimize the features so that a feature was not divided by a boundary. Mild smoothing was included (50 point smoothing) since the stellar features are broad. These velocity differences are directly used in our orbital model fitting. The uncertainties were derived from the differences between segments. Measurements are listed in Table~\ref{table_rv2}.

	\begin{table}[h]
		\centering
		\caption{Differential radial velocities of the companion of V1334~Cyg.} 
		\begin{tabular}{cc}
			\hline
			\hline
			MJD  & $\Delta V_2$ \\
			& 	(km~s$^{-1}$)						    \\
			\hline
			56916.757  &   $0.00\pm0.51$ \\
			56918.548  &   $0.47\pm0.43$ \\
			57251.901  &   $-8.70\pm0.39$ \\
			57645.272  &   $-28.36\pm0.71$ \\
			\hline
		\end{tabular}
		\label{table_rv2}
	\end{table}

\section{Model fitting and orbital solutions}
\label{section__model_fitting_and_model_fitting}

	We performed a combined fit of the interferometric orbit and RVs of the Cepheid and its companion. The fit includes three models with shared parameters for our three datasets. The first models the Cepheid radial velocities only (SOPHIE and HERMES data), i.e. the orbit around the system barycentre and the pulsation. The second models the companion's relative radial velocities (STIS data), i.e. the orbit of the secondary around the system barycentre. The last model fits the relative astrometric orbit of the companion determined from interferometry (MIRC data).

	Our radial velocity model of the primary star, the Cepheid, is defined as
	\begin{displaymath}
		V_1(t) = V_\mathrm{1,orb} + V_\mathrm{puls},
	\end{displaymath}
	with the orbital radial velocity $V_\mathrm{1,orb}$ and the pulsation velocity $V_\mathrm{puls}$, which are expressed with
	\begin{eqnarray*}
		V_\mathrm{1,orb} &=& K_1 [\cos(\omega + \nu) + e \cos \omega] + v_\gamma, \\
		V_\mathrm{puls} &=& \sum_{i=1}^n [A_i \cos(2\pi i \phi_\mathrm{puls}) + B_i \sin(2\pi i \phi_\mathrm{puls})],
	\end{eqnarray*}
	with $K_1$ the semi-amplitude of the Cepheid’s orbit due to the companion, $\omega$ the argument of periastron of the companion’s orbit, $\nu$ the true anomaly of the companion, $e$ the eccentricity of the orbit, $v_\gamma$ the systemic velocity, the pulsation phase $\phi_\mathrm{puls} = (t - T_0)/P_\mathrm{puls}$ (modulo 1), and $(A_i, B_i)$ the amplitude of the Fourier series. The true anomaly is defined implicitly with the following three Keplerian parameters $P_\mathrm{orb} , T_p$ and $e$ and Kepler's equation:
	\begin{eqnarray*}
		\tan \frac{\nu (t)}{2} &=& \sqrt{\frac{1 + e}{1 - e} } \tan \frac{E(t)}{2}\\
		E(t) - e \sin E(t) &=& \frac{2\pi (t - T_p)}{P_\mathrm{orb}},
	\end{eqnarray*}
	where $E(t)$ is the eccentric anomaly, $T_p$ the time of periastron passage, $t$ the time of radial velocity observations, and $P_\mathrm{orb}$ the orbital period.

	Following the linear parametrization developed in \citet{Wright_2009_05_0} for $V_\mathrm{1,orb}$, and including now the pulsation, our model can be simplified to
	\begin{equation}
		\begin{split}
			V_1&(t) = C_1 \cos \nu + C_2 \sin \nu + v_0\\
			&+ \sum_{i=1}^2 [A_i \cos(2\pi i \phi_\mathrm{puls}) + B_i \sin(2\pi i \phi_\mathrm{puls})
		\end{split}
	\end{equation}
	where we restricted the Fourier series to $n = 2$ as the pulsation curve is sinusoidal-like for such short-period Cepheid (see also Fig.~\ref{figure_lightcurve}). The parameters $C_1, C_2$ and $v_0$ are related to the Keplerian parameters through the relations \citep{Wright_2009_05_0}:
	\begin{eqnarray*}
		C_1 &=& K_1 \cos \omega, \\
		C_2 &=& -K_1 \sin \omega, \\
		v_0 &=& v_\gamma + K_1 e \cos \omega,
	\end{eqnarray*}
	which can be converted back with ($\omega$ chosen so that $\sin \omega$ has the sign of the numerator):
	\begin{eqnarray*}
		K_1 &=& \sqrt{C_1^2 + C_2^2}, \\
		\tan w &=& \frac{-C_2}{C_1},\\
		v_\gamma &=& v_0 - K_1 e \cos \omega.
	\end{eqnarray*}
	The fitted parameters are therefore defined as ($C_1, C_2$, $v_0, A_1, B_1, A_2, B_2, P_\mathrm{orb}, T_p, e, P_\mathrm{puls}$). $T_0$ is kept fixed in the fitting process to avoid degeneracy with $T_p$. The value is chosen such that $V_\mathrm{puls}(0) = 0$.

	For the second model, as we cross-correlated each STIS spectra with respect to the first observation, our relative radial velocities of the companion can be parametrized with:

	\begin{eqnarray}
		\Delta V_\mathrm{2,orb}(t) &=& V_\mathrm{2,orb}(t) - V_\mathrm{2,orb}(t_0), \\
		V_\mathrm{2,orb}(t) &=& \frac{K_1}{q} [\cos(\omega + \nu) + e \cos \omega] + v_\gamma
	\end{eqnarray}
	with the mass ratio $q = M_2/M_1$, and $t_0$ the time of the first STIS measurement. This adds the new parameter, $q$ as fitted parameter.

	Finally, the astrometric positions of the companion as measured from interferometry are modelled with the following equations \citep{Heintz_1978__0}:
	\begin{eqnarray}
		\Delta \alpha &=& B X + G Y, \\
		\Delta \delta &=& A X + F Y, 
	\end{eqnarray}
	where $(X,Y)$ are the elliptical rectangular coordinates defined as:
	\begin{eqnarray*}
		X(t) &=& \cos E(t) - e \\
		Y(t) &=& \sqrt{1 - e^2} \sin E(t)
	\end{eqnarray*}

	\begin{figure*}[ht]
		\centering
		\resizebox{\hsize}{!}{\includegraphics{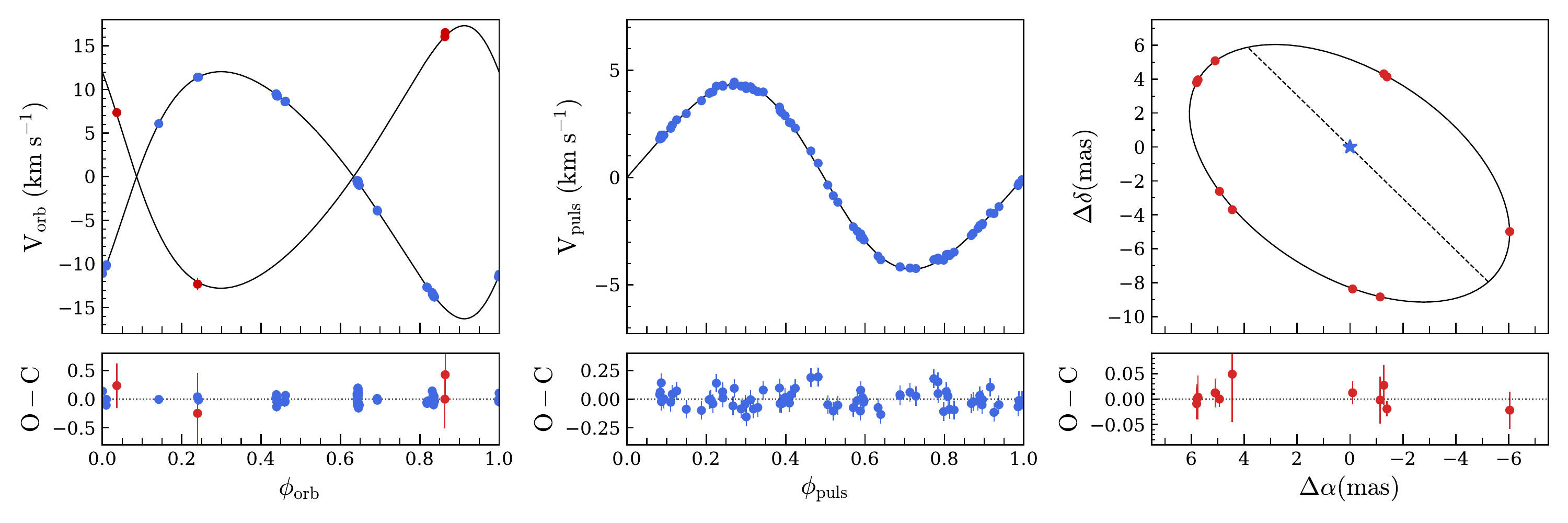}}
		\caption{Result of our combined fit. Left: fitted (solid lines) and extracted primary (blue dots) and secondary (red dots) orbital velocity. Middle: fitted (solid line) and extracted (blue dots) pulsation velocity. Right: relative astrometric orbit of V1334 Cyg Ab.}
		\label{figure_orbit}
	\end{figure*}

	The Thiele-Innes constants are parametrized in term of the orbital elements with:
	\begin{eqnarray*}
		A &=&  a_\mathrm{as} (\cos \Omega \cos \omega - \sin \Omega \sin \omega \cos i), \\
		B &=& a_\mathrm{as} (\sin \Omega \cos \omega + \cos \Omega \sin \omega \cos i), \\
		F &=& a_\mathrm{as} (-\cos \Omega \sin \omega - \sin \Omega \cos \omega \cos i), \\
		G &=&  a_\mathrm{as} (-\sin \Omega \sin \omega + \cos \Omega \cos \omega \cos i),
	\end{eqnarray*}
	with $a_\mathrm{as}$ the angular semi-major axis in arcsecond (as), $\Omega$ the position angle of the ascending node, and $i$ the orbital inclination, which are determined through the inversion relations:
	\begin{eqnarray*}
		\tan (\Omega + \omega) &=& \frac{B-F}{A + G}, \\
		\tan (\Omega - \omega) &=& \frac{B+F}{A - G}, \\
		k_1 &=& \frac{1}{2} (A^2 + B^2 + F^2 + G^2), \\
		k_2 &=& AG - BF, \\
		a_\mathrm{as}^2 &=& \sqrt{k_1^2 - k_2^2} + k_,1 \\
		\cos i &=& \frac{k_2}{a_\mathrm{as}^2},
	\end{eqnarray*}
	with the quadrant of $\Omega \pm \omega$ determined from the signs of the numerators. These Thiele-Innes constants are included to the parameter list, giving a total of 16 variables to fit.

	The distance and masses are then given by:
	\begin{eqnarray*}
		K_2 &=& \frac{K_1}{q}, \\
		a_\mathrm{au} &=&  \frac{9.191966\times10^{-5} (K_1 + K_2) P_\mathrm{orb} \sqrt{1 - e^2}}{\sin^3 i}, \\
		d &=& \frac{a_\mathrm{au}}{a_\mathrm{as}}, \\
		M_T &=& M_1 + M_2 = \frac{a^3 d^3}{P_\mathrm{orb}^2}, \\
		M_1 &=& \frac{M_T}{1 + q}, \\
		M_2 &=& q\,M_1,
	\end{eqnarray*}
	with $K_2$ the semi-amplitude of the companion, $a_\mathrm{au}$ the linear semi-major axis in astronomical unit (au), $d$ the distance to the system, and $M_1$ and $M_2$ the Cepheid and companion mass, respectively.

	We then applied a classical Monte Carlo Markov Chain (MCMC) technique to fit the 16 model parameters characterizing the standard orbital elements and the pulsation of the Cepheid. Our best fit values were obtained from the median values of the MCMC distribution (100~000 samples with uniform priors), while uncertainties were derived from the maximum value between the 16\,\% and 84\,\% percentiles. Our best fit is shown in Fig.~\ref{figure_orbit}. We determined a distance $d = \dist \pm \edist$\,pc ($\pm 1.1$\,\%), which is the most accurate model-independent distance of a Cepheid. This corresponds to a parallax of $\pi = \parallax \pm \eparallax$\,mas. The two component masses were also determined with a high accuracy: $M_1 = \Mcep \pm \eMcep\,M_\odot$ ($\pm 3.1$\,\%) and $M_2 = \Mcomp \pm \eMcomp\,M_\odot$ ($\pm 1.2$\,\%), with the primary star being the Cepheid. The other derived orbital and pulsation parameters are listed in Table~\ref{table_orbit}.
	
	Note that our orbit solutions are rather good agreement with the previous estimates of the spectroscopic elements determined by \citet[][also listed in Table~\ref{table_orbit}]{Evans_2000_06_0}. It is worth mentioning that we did not use additional RVs measurements from the literature for several reasons: 1) they are usually not very precise, 2) we wanted to use a dataset as uniform as possible (i.e. RVs estimated the same way), 3) the effect on the RVs of a possible third component is reduced, and 4) we also avoid possible bias from the changing pulsation period of the Cepheid by limiting the time range.


		\begin{table}[!h]
	\centering
	\caption{Final estimated parameters of the V1334~Cyg system. Index 1 designates the Cepheid and index 2 its main sequence companion. Note that $T_0$ is kept fixed.}
	\begin{tabular}{ccc} 
		\hline
		\hline
		\multicolumn{3}{c}{Pulsation}																	\\
		$P_\mathrm{puls}$ (days)							  &  \multicolumn{2}{c}{$3.33242 \pm 0.00002$}\\
		$T_0$ (JD)						   						 	 &  \multicolumn{2}{c}{$2~445~000.550$}		\\
		$A_1$ (km~s$^{-1}$)					&	\multicolumn{2}{c}{$-0.20 \pm 0.35$}			\\
		$B_1$ (km~s$^{-1}$)					&	\multicolumn{2}{c}{$4.22 \pm 0.04$}		\\
		$A_2$ (km~s$^{-1}$)					&	\multicolumn{2}{c}{$-0.01 \pm 0.08$}		\\
		$B_2$ (km~s$^{-1}$)					&	\multicolumn{2}{c}{$-0.42 \pm 0.01$}		\\
		\hline
		\multicolumn{3}{c}{Orbit}  																			\\
																					&   Evans+ (2000) &	This work  			\\
			$P_\mathrm{orb}$ (days)									&  $1937.5 \pm 2.1$ & $1932.8 \pm 1.8$ 		  \\
$T_\mathrm{p}$ (JD)										& $2443607 \pm 14$  & $2453316.75 \pm 4.1$   \\
$e$																     & $0.197 \pm 0.009$ &  $0.233 \pm 0.001$	\\
$\omega$	($^\circ$)									 &  $226.3 \pm 2.9$  &	$229.8 \pm 0.3$		  \\
$K_1$ ($\mathrm{km~s^{-1}}$)					  & $14.1 \pm 0.1$	&	$14.168 \pm 0.014$	\\
$K_2$ ($\mathrm{km~s^{-1}}$)					  &	--	&	$15.036 \pm 0.304$	\\
$v_\gamma$	($\mathrm{km~s^{-1}}$)			&  $-1.8 \pm 0.1$	&	$-2.65 \pm 0.01$		\\
$\Omega$	($^\circ$)									 &	--	&	$213.17 \pm 0.35$	\\
$i$ ($^\circ$)													&	--	&	$124.94 \pm 0.09$	\\
$a$ (mas)														 &	--	&	$8.54 \pm 0.04$	\\
$a$ (au)														 &	--	&	$6.16 \pm 0.07$	 \\
$d$ (pc)														 &	--	&	$\dist \pm \edist$	\\
$q$ 																&	--	&	$\Mratio \pm \eMratio$	\\
$M_1$ ($M_\odot$)											&	--	&	$\Mcep \pm \eMcep$	\\
$M_2$ ($M_\odot$)											&	--	&	$\Mcomp \pm \eMcomp$	\\
		\hline															
	\end{tabular}
	\label{table_orbit}
\end{table}

\section{Discussion and conclusions}
\label{section__discussion_and_conclusions}

	This is the first time that a binary Cepheid is resolved both spatially and spectroscopically. This provides the most accurate model-independent distance of a Cepheid, and the most precise masses for a Galactic Cepheid. Here we do not discuss or compare the evolutionary status of this system using our mass measurements as this can be found in \citet{Evans_2018_08_1}.
	
	With such a high distance accuracy, it is possible to test the agreement of V1334 Cyg with the existing calibrations of the P--L relations. We selected a sample of published relations in the $V$- and $K$ filters, as they are the most frequently used photometric bands \citep{Sandage_2004_09_0,Fouque_2007_12_0,Benedict_2007_04_0, Storm_2011_10_0,Groenewegen_2013_02_0,Gaia-Collaboration_2017_09_0}.

	This Cepheid is classified as pulsating in its first overtone, with a slightly shorter period than in the fundamental mode. We therefore converted the period from the fundamental to the first overtone mode following the method used in \citet{Pilecki_2018_07_0} with a pulsation model, and described in \citet{Taormina_2018_03_0}, resulting in an equivalent fundamental period of $P_0 = 4.7897\pm0.0322$\,days. The comparison with the empirical relation from \citet{Alcock_1995_04_0}, $P_1/P_0 = 0.720 - 0.027 \log{P_0}$, is very good with a relative difference $< 1\,$\% (or within $1.3\sigma$).

	\begin{figure*}[!ht]
		\centering
		\resizebox{\hsize}{!}{\includegraphics{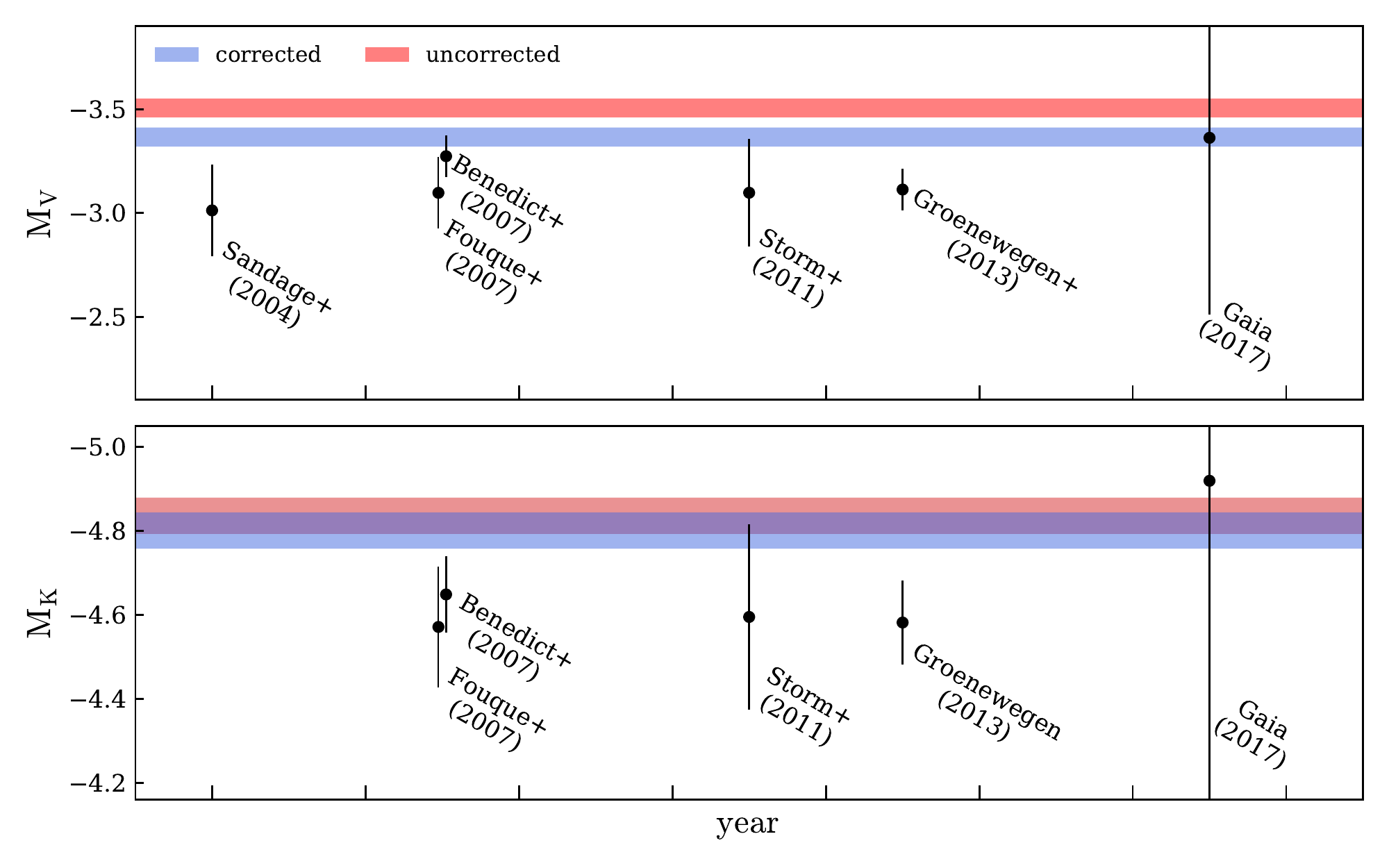}}
		\caption{Comparison between the absolute magnitudes of V1334~Cyg predicted from literature P-L relations (black dots) and the present distance measurement (red and blue areas), in two photometric bands. The red and blue areas represent the measured absolute magnitude without and with the subtraction of the companion's flux contamination, respectively. In the $K$ band, these two regions overlap as the contribution of the companion is smaller.}
		\label{figure_pl}
	\end{figure*}

	\begin{figure}[!ht]
		\centering
		\resizebox{\hsize}{!}{\includegraphics{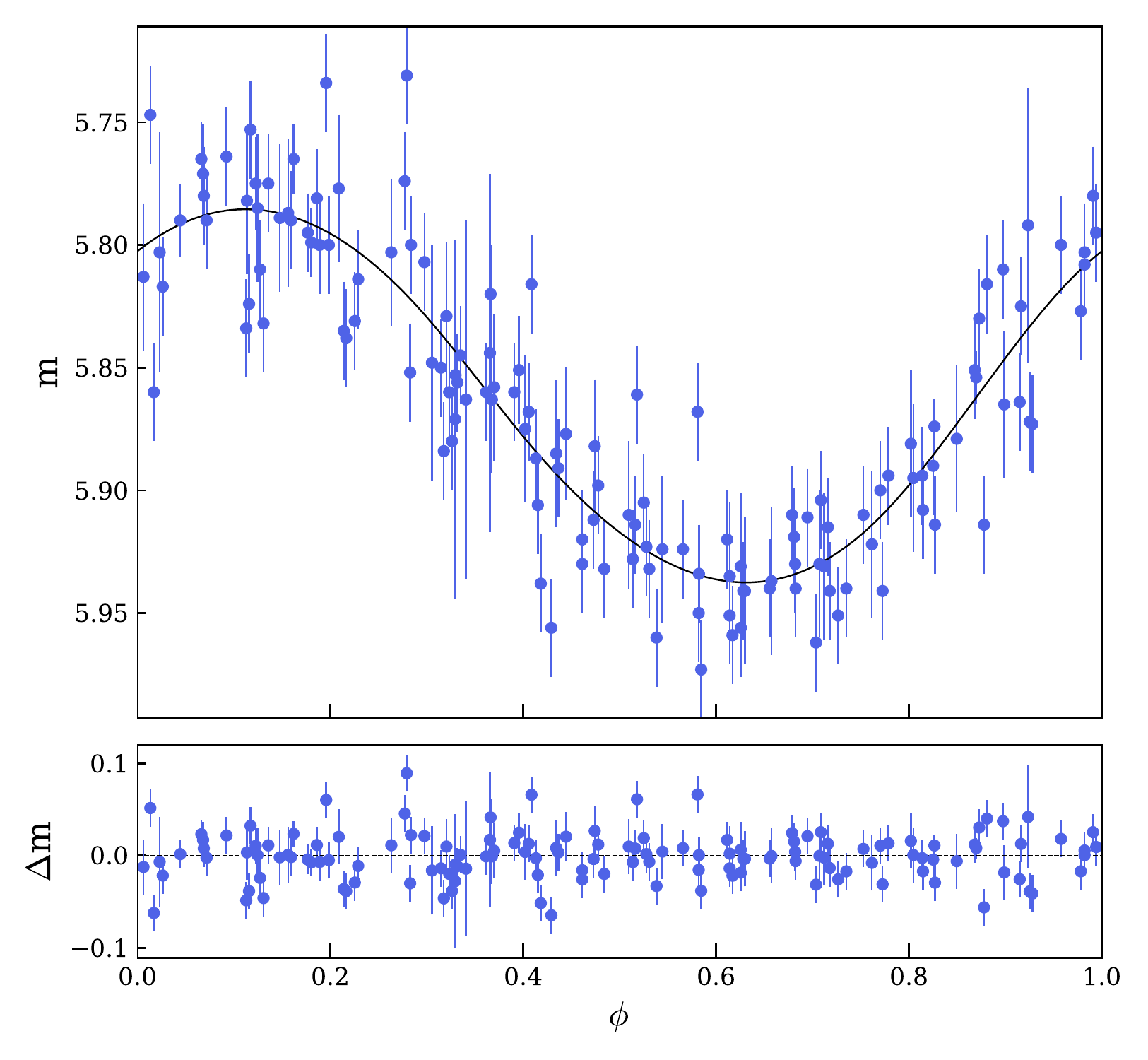}}
		\caption{Interpolated $V$-band light curve of V1334~Cyg.}
		\label{figure_lightcurve}
	\end{figure}
	\begin{figure}[!ht]
	\centering
	\resizebox{\hsize}{!}{\includegraphics{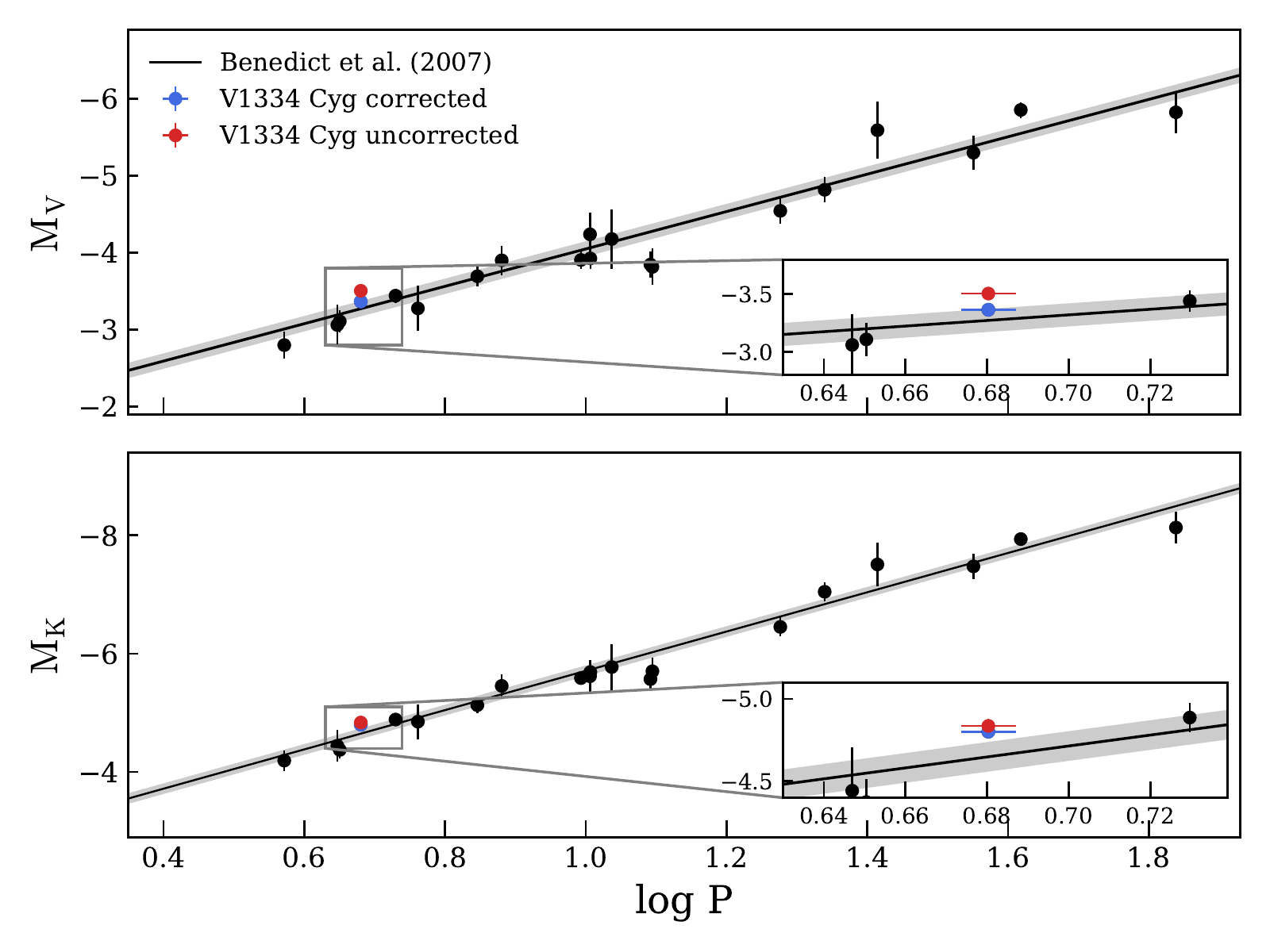}}
	\caption{Period-Luminosity relations constructed from direct parallax measurements. This includes our measured distance and those from \cite{Casertano_2016_07_0,Riess_2018_03_0,Benedict_2007_04_0,Benedict_2002_09_0}. We also included the distance estimated from light echoes by \cite{Kervella_2014_12_0}. The blue and red dot represent the location of V1334~Cyg for companion-flux corrected and uncorrected magnitude, respectively. Uncertainty on the "fundamentalized" period is also shown. The relation derived by \cite{Benedict_2007_04_0} is also plotted, with the rms of the residuals in shaded area.}
	\label{figure_PL2}
\end{figure}

	For each tested P--L relation, we determined the predicted absolute magnitudes for $P_0$, that are represented in Fig.~\ref{figure_pl} as black dots. We adopted as uncertainties the scatter of each relation given by the authors, which better represents the observed intrinsic dispersion \citep[the width of the instability strip is the dominating uncertainty in the P-L relations, see e.g][for a recent study]{Anderson_2016_06_0}. We derived the absolute magnitude with the usual relation $M_\lambda = m_\lambda - A_\lambda - 5\,\log d + 5$, with $m_\lambda$ the apparent magnitude at wavelength $\lambda$, $A_\lambda$ the interstellar absorption parameter, and $d$ the distance to the star. We corrected for the interstellar extinction, using $A_\mathrm{V} = 3.23 E(B-V)$ and $A_\mathrm{K} = 0.119 A_\mathrm{V}$ \citep{Fouque_2007_12_0}, with $E(B-V) = 0.025 \pm 0.009$ \citep{Kovtyukh_2008_09_0}. For the $V$ band, we estimated the weighted mean apparent magnitude using a periodic cubic spline fit of four different published light curves \citep{Henden_1980_08_0,Berdnikov_2008_04_0,Szabados_1977_01_0,Kiss_1998_07_0}. The curve is shown in Fig.~\ref{figure_lightcurve}. We determined the weighted mean dereddenned magnitude $m_\mathrm{0,V} = 5.781 \pm 0.026$\,mag, where the total uncertainty is estimated from the standard deviation of the residual values. There are no light curves available in the near-infrared, so to estimate $m_\mathrm{0,K}$ we used the value measured by 2MASS \citep{Cutri_2003_03_0}. These observations were obtained at the pulsation phase 0.36, very close to the phase of the mean radius (0.37), so that the magnitude variation is negligible (within the uncertainty). We estimated $m_\mathrm{0,K} = 4.451 \pm 0.036$\,mag for the mean apparent magnitude in the $K$ band. We then combined with our measured distance to estimate the absolute magnitudes $M_\lambda$, for which the total error bar includes the uncertainties on $d, A_\lambda$, and $m_\lambda$. Fig.~\ref{figure_pl} shows the difference between $M_\lambda$ of V1334 Cyg (red area) and the predicted values from literature P--L relations (black dots). However, the contribution of the companion must be subtracted from the combined magnitudes. To correct from the flux contamination of the companion, we used the magnitude difference between the components estimated in $V$ by \cite{Evans_1995_05_0}, $\Delta V = 2.18$\,mag, and from our measured $H$-band average flux ratio from interferometry for the $K$ band, i.e. we assumed $\Delta K = \Delta H = 3.70 \pm 0.11$\,mag. With our distance we can estimate the mean corrected absolute magnitudes of the Cepheid to be $M\mathrm{_V(cep)} = -3.37 \pm 0.05$\,mag, $M\mathrm{_H(cep)} = -4.60 \pm 0.05$\,mag and $M\mathrm{_K(cep)} = -4.80 \pm 0.04$\,mag. Note that this is a rough estimate in $K$ as we assumed the flux ratio to be the same as in $H$. From the magnitude differences, absolute magnitudes for the companion can also be derived, we found $M\mathrm{_V(comp)} = -1.19 \pm 0.11$\,mag, $M\mathrm{_H(comp)} = -0.90 \pm 0.12$\,mag and $M\mathrm{_K(comp)} = -1.10 \pm 0.12$\,mag. The comparison of the $(V - H)_0$ colour with the intrinsic colour table of \citet{Pecaut_2013_09_0}\footnote{see also \url{http://www.pas.rochester.edu/~emamajek/EEM_dwarf_UBVIJHK_colors_Teff.txt}} corresponds to a $\sim$B7V companion of $\sim 3.9\,M_\odot$ (the $(V - K)_0$ colour is more uncertain as we assumed the same flux ratio in $K$ and $H$). This is in good agreeement with the spectral type estimate of \citet[][mean spectrum of a B7V star]{Evans_1995_05_0} and consistent with our measured mass.
	
	The blue area in Fig.~\ref{figure_pl} shows the absolute magnitudes of the Cepheid corrected from the flux contamination. We see in the optical band that the relations based on trigonometric parallax measurements \citep{Benedict_2007_04_0,Gaia-Collaboration_2017_09_0} are in agreement with the measured corrected absolute magnitude of the Cepheid. In the infrared, the agreement is better, but still points to an offset, and again direct distance estimates seem to be more consistent. This can be explained by the fact that photometric measurements are more sensitive to a companion's flux contamination than are parallax observations, and therefore more strongly biased. The relation from \cite{Storm_2011_10_0} is in marginal agreement, but also has a large uncertainty. In general, we see a systematic offset in P-L relations derived through photometric measurements (Baade-Wesselink method). In $V$, the average offset for photometric-based absolute magnitudes only is $\sim 1.5\sigma$, and $\sim 1.2\sigma$ for the $K$ band. It is worth mentioning that increasing the value of $E(B-V)$ would have the effect of increasing the inconsistency of our estimated absolute magnitude and the ones from the P-L relations plotted in Fig.~\ref{figure_pl}. We will also note that as V1334~Cyg is a first overtone pulsator, therefore the amplitude in temperature can be neglected \citep[$\sim 240$\,K,][]{Luck_2008_07_0}.

	V1334~Cyg is a clear demonstration that Cepheid companions in binary systems are a significant source of a systematically positive photometric bias to the Cepheid fluxes, as well as contributors to the scattering of the P--L relations. In Fig.~\ref{figure_PL2} we show that correcting for the companion's flux brings the star closer to the position predicted by the P--L relations, with a stronger effect in $V$ for this specific system. The same effect has been observed in the $V$ and $I$ band by \citet{Pilecki_2018_07_0} in six LMC Cepheids in eclipsing binary systems.
    
    Finally, the comparison of our orbital parallax with the second Gaia data release \citep[$\pi = 1.151\pm0.066\,\mathrm{mas}$][]{Gaia-Collaboration_2018_08_0} shows a $3.6\sigma$ deviation with our value. Correcting for the Gaia systematic offset of $\sim -0.029$\,mas \citep[estimated for quasars,][]{Lindegren_2018_08_0} still leaves a disagreement at $3.2\sigma$. This is however not surprising as binarity is not yet taken into account in the Gaia data reduction process. V1334 Cyg stands out as a demonstration that binarity can significantly impact Cepheid parallaxes. The selection criteria of \citet{Gaia-Collaboration_2017_09_0} of rejecting the known binary Cepheids seems a good approach, as seen in our Fig.~\ref{figure_pl} with a good agreement in $V$ band between the expected value and our measured corrected absolute magnitude. The final Gaia data release with hundreds of Galactic Cepheid parallaxes promises to constrain the zero point of P--L relations to an exquisite precision. But reddening corrections, biases from orbiting companions (astrometric and photometric), as well as the flux contribution of circumstellar envelopes will limit the accuracy of the P-L relation calibration if not properly taken into account.

\acknowledgments

This research is based on observations made with SOPHIE spectrograph on the 1.93-m telescope at Observatoire de Haute-Provence (CNRS/AMU), France (ProgID: 13A.PNPS10, 13B.PNPS003, 14A.PNPS010, 15A.PNPS010, 16B.PNPS.KERV). This research is based on observations made with the Mercator Telescope, operated on the island of La Palma by the Flemish Community, at the Spanish Observatorio del Roque de los Muchachos of the Instituto de Astrofsica de Canarias. Hermes is supported by the Fund for Scientific Research of Flanders (FWO), Belgium; the Research Council of K.U.Leuven, Belgium; the Fonds National de la Recherche Scientifique (F.R.S.- FNRS), Belgium; the Royal Observatory of Belgium; the Observatoire de Genève, Switzerland; and the Th\"{u}ringer Landessternwarte, Tautenburg, Germany. This work is also based on observations with the NASA/ESA Hubble Space Telescope obtained at the Space Telescope Science Institute, which is operated by the Association of Universities for Research in Astronomy, Inc., under NASA contract NAS5-26555 (ProgID 13454). We acknowledge the support of the French Agence Nationale de la Recherche (ANR-15-CE31-0012-01, project Unlock-Cepheids). WG and GP gratefully acknowledge financial support from the BASAL Centro de Astrofisica y Tecnologias Afines (CATA, AFB-170002). WG also acknowledges financial support from the Millenium Institute of Astrophysics (MAS) of the Iniciativa Cientifica Milenio del Ministerio de Economia, Fomento y Turismo de Chile (project IC120009). We acknowledge financial support from the Programme National de Physique Stellaire (PNPS) of CNRS/INSU, France. Support from the Polish National Science Centre grants MAESTRO UMO-2017/26/A/ST9/00446 and from the IdP II 2015 0002 64 grant of the Polish Ministry of Science and Higher Education is also acknowledged. The research leading to these results has received funding from the European Research Council (ERC) under the European Union's Horizon 2020 research and innovation programme (grant agreement No. 695099 and 639889). NRE acknowledge support from the Chandra X-ray Center NASA (contract NAS8-03060) and the HST grants GO-13454.001-A and  GO-14194.002. This work is based upon observations obtained with the Georgia State University Center for High Angular Resolution Astronomy Array at Mount Wilson Observatory. The CHARA Array is supported by the National Science Foundation under Grants No. AST-1211929, 1411654, and 1636624. Institutional support has been provided from the GSU College of Arts and Sciences and the GSU Office of the Vice President for Research and Economic Development. BP acknowledges financial support from the Polish National Science Center grant SONATA 2014/15/D/ST9/02248.

%

\vspace{5mm}
\facilities{HST(STIS), CHARA(MIRC), OHP:1.93m (SOPHIE), La Palma:1.2m (HERMES)}


\software{Candid \citep{Gallenne_2015_07_0}, Emcee \citep{Foreman-Mackey_2013_03_0}, SearchCal \citep{Bonneau_2006_09_0}, MIRC pipeline \citep{Monnier_2007_07_0}}





\bibliographystyle{aasjournal}   
\bibliography{/Users/agallenn/Sciences/Articles/bibliographie}

\begin{thebibliography}{}
\expandafter\ifx\csname natexlab\endcsname\relax\def\natexlab#1{#1}\fi

\bibitem[{{Alcock} {et~al.}(1995){Alcock}, {Allsman}, {Axelrod}, {Bennett},
  {Cook}, {Freeman}, {Griest}, {Marshall}, {Peterson}, {Pratt}, {Quinn},
  {Reimann}, {Rodgers}, {Stubbs}, {Sutherland}, \& {Welch}}]{Alcock_1995_04_0}
{Alcock}, C., {Allsman}, R.~A., {Axelrod}, T.~S., {et~al.} 1995, \aj, 109, 1653

\bibitem[{{Anderson} {et~al.}(2016{\natexlab{a}}){Anderson}, {Saio},
  {Ekstr{\"o}m}, {Georgy}, \& {Meynet}}]{Anderson_2016_06_0}
{Anderson}, R.~I., {Saio}, H., {Ekstr{\"o}m}, S., {Georgy}, C., \& {Meynet}, G.
  2016{\natexlab{a}}, \aap, 591, A8

\bibitem[{{Anderson} {et~al.}(2016{\natexlab{b}}){Anderson}, {Casertano},
  {Riess}, {Melis}, {Holl}, {Semaan}, {Papics}, {Blanco-Cuaresma}, {Eyer},
  {Mowlavi}, {Palaversa}, \& {Roelens}}]{Anderson_2016_10_0}
{Anderson}, R.~I., {Casertano}, S., {Riess}, A.~G., {et~al.}
  2016{\natexlab{b}}, \apjs, 226, 18

\bibitem[{{Benedict} {et~al.}(2002){Benedict}, {McArthur}, {Fredrick},
  {Harrison}, {Slesnick}, {Rhee}, {Patterson}, {Skrutskie}, {Franz},
  {Wasserman}, {Jefferys}, {Nelan}, {van Altena}, {Shelus}, {Hemenway},
  {Duncombe}, {Story}, {Whipple}, \& {Bradley}}]{Benedict_2002_09_0}
{Benedict}, G.~F., {McArthur}, B.~E., {Fredrick}, L.~W., {et~al.} 2002, \aj,
  124, 1695

\bibitem[{{Benedict} {et~al.}(2007){Benedict}, {McArthur}, {Feast}, {Barnes},
  {Harrison}, {Patterson}, {Menzies}, {Bean}, \&
  {Freedman}}]{Benedict_2007_04_0}
{Benedict}, G.~F., {McArthur}, B.~E., {Feast}, M.~W., {et~al.} 2007, \aj, 133,
  1810

\bibitem[{{Berdnikov}(2008)}]{Berdnikov_2008_04_0}
{Berdnikov}, L.~N. 2008, VizieR Online Data Catalog: II/285, originally
  published in: Sternberg Astronomical Institute, Moscow, 2285

\bibitem[{{Bohm-Vitense}(1985)}]{Bohm-Vitense_1985_09_1}
{Bohm-Vitense}, E. 1985, \apj, 296, 169

\bibitem[{{Bonneau} {et~al.}(2006){Bonneau}, {Clausse}, {Delfosse}, {Mourard},
  {Cetre}, {Chelli}, {Cruzal{\`e}bes}, {Duvert}, \& {Zins}}]{Bonneau_2006_09_0}
{Bonneau}, D., {Clausse}, J.-M., {Delfosse}, X., {et~al.} 2006, \aap, 456, 789

\bibitem[{{Bouchy} {et~al.}(2009){Bouchy}, {H{\'e}brard}, {Udry}, {Delfosse},
  {Boisse}, {Desort}, {Bonfils}, {Eggenberger}, {Ehrenreich}, {Forveille},
  {Lagrange}, {Le Coroller}, {Lovis}, {Moutou}, {Pepe}, {Perrier}, {Pont},
  {Queloz}, {Santos}, {S{\'e}gransan}, \& {Vidal-Madjar}}]{Bouchy_2009_10_0}
{Bouchy}, F., {H{\'e}brard}, G., {Udry}, S., {et~al.} 2009, \aap, 505, 853

\bibitem[{{Casertano} {et~al.}(2016){Casertano}, {Riess}, {Anderson},
  {Anderson}, {Bowers}, {Clubb}, {Cukierman}, {Filippenko}, {Graham},
  {MacKenty}, {Melis}, {Tucker}, \& {Upadhya}}]{Casertano_2016_07_0}
{Casertano}, S., {Riess}, A.~G., {Anderson}, J., {et~al.} 2016, \apj, 825, 11

\bibitem[{{Cutri} {et~al.}(2003){Cutri}, {Skrutskie}, {van Dyk}, {Beichman},
  {Carpenter}, {Chester}, {Cambresy}, {Evans}, {Fowler}, {Gizis}, {Howard},
  {Huchra}, {Jarrett}, {Kopan}, {Kirkpatrick}, {Light}, {Marsh}, {McCallon},
  {Schneider}, {Stiening}, {Sykes}, {Weinberg}, {Wheaton}, {Wheelock}, \&
  {Zacarias}}]{Cutri_2003_03_0}
{Cutri}, R.~M., {Skrutskie}, M.~F., {van Dyk}, S., {et~al.} 2003, VizieR Online
  Data Catalog, 2246, 0

\bibitem[{{Evans}(1992)}]{Evans_1992_01_0}
{Evans}, N.~R. 1992, \apj, 384, 220

\bibitem[{{Evans}(1995)}]{Evans_1995_05_0}
---. 1995, \apj, 445, 393

\bibitem[{{Evans}(2000)}]{Evans_2000_06_0}
---. 2000, \aj, 119, 3050

\bibitem[{{Evans} {et~al.}(2018{\natexlab{a}}){Evans}, {Proffitt}, {Carpenter},
  {Winston}, {Kober}, {G{\"u}nther}, {Gorynya}, {Rastorguev}, \&
  {Inno}}]{Evans_2018_09_0}
{Evans}, N.~R., {Proffitt}, C., {Carpenter}, K.~G., {et~al.}
  2018{\natexlab{a}}, \apj, in press [e-prints: 1808.10472]

\bibitem[{{Evans} {et~al.}(2018{\natexlab{b}}){Evans}, {Karovska}, {Bond},
  {Schaefer}, {Sahu}, {Mack}, {Nelan}, {Gallenne}, \&
  {Tingle}}]{Evans_2018_08_1}
{Evans}, N.~R., {Karovska}, M., {Bond}, H.~E., {et~al.} 2018{\natexlab{b}},
  \apj, 863, 187

\bibitem[{{Foreman-Mackey} {et~al.}(2013){Foreman-Mackey}, {Hogg}, {Lang}, \&
  {Goodman}}]{Foreman-Mackey_2013_03_0}
{Foreman-Mackey}, D., {Hogg}, D.~W., {Lang}, D., \& {Goodman}, J. 2013, \pasp,
  125, 306

\bibitem[{{Fouqu{\'e}} {et~al.}(2007){Fouqu{\'e}}, {Arriagada}, {Storm},
  {Barnes}, {Nardetto}, {M{\'e}rand}, {Kervella}, {Gieren}, {Bersier},
  {Benedict}, \& {McArthur}}]{Fouque_2007_12_0}
{Fouqu{\'e}}, P., {Arriagada}, P., {Storm}, J., {et~al.} 2007, \aap, 476, 73

\bibitem[{{Freedman} \& {Madore}(2010)}]{Freedman_2010_09_0}
{Freedman}, W.~L., \& {Madore}, B.~F. 2010, \araa, 48, 673

\bibitem[{{Gaia Collaboration} {et~al.}(2017){Gaia Collaboration},
  {Clementini}, {Eyer}, {Ripepi}, {Marconi}, {Muraveva}, {Garofalo}, {Sarro},
  {Palmer}, {Luri}, \& et~al.}]{Gaia-Collaboration_2017_09_0}
{Gaia Collaboration}, {Clementini}, G., {Eyer}, L., {et~al.} 2017, \aap, 605,
  A79

\bibitem[{{Gaia Collaboration} {et~al.}(2018){Gaia Collaboration}, {Brown},
  {Vallenari}, {Prusti}, {de Bruijne}, {Babusiaux}, {Bailer-Jones}, {Biermann},
  {Evans}, {Eyer}, \& et~al.}]{Gaia-Collaboration_2018_08_0}
{Gaia Collaboration}, {Brown}, A.~G.~A., {Vallenari}, A., {et~al.} 2018, \aap,
  616, A1

\bibitem[{{Gallenne} {et~al.}(2017){Gallenne}, {Kervella}, {M{\'e}rand},
  {Pietrzy{\'n}ski}, {Gieren}, {Nardetto}, \& {Trahin}}]{Gallenne_2017_11_0}
{Gallenne}, A., {Kervella}, P., {M{\'e}rand}, A., {et~al.} 2017, \aap, 608, A18

\bibitem[{{Gallenne} {et~al.}(2013{\natexlab{a}}){Gallenne}, {M{\'e}rand},
  {Kervella}, {Chesneau}, {Breitfelder}, \& {Gieren}}]{Gallenne_2013_10_0}
{Gallenne}, A., {M{\'e}rand}, A., {Kervella}, P., {et~al.} 2013{\natexlab{a}},
  \aap, 558, A140

\bibitem[{{Gallenne} {et~al.}(2013{\natexlab{b}}){Gallenne}, {Monnier},
  {M{\'e}rand}, {Kervella}, {Kraus}, {Schaefer}, {Gieren}, {Pietrzy{\'n}ski},
  {Szabados}, {Che}, {Baron}, {Pedretti}, {McAlister}, {ten Brummelaar},
  {Sturmann}, {Sturmann}, {Turner}, {Farrington}, \&
  {Vargas}}]{Gallenne_2013_04_0}
{Gallenne}, A., {Monnier}, J.~D., {M{\'e}rand}, A., {et~al.}
  2013{\natexlab{b}}, \aap, 552, A21

\bibitem[{{Gallenne} {et~al.}(2015){Gallenne}, {M{\'e}rand}, {Kervella},
  {Monnier}, {Schaefer}, {Baron}, {Breitfelder}, {Le Bouquin}, {Roettenbacher},
  {Gieren}, {Pietrzy{\'n}ski}, {McAlister}, {ten Brummelaar}, {Sturmann},
  {Sturmann}, {Turner}, {Ridgway}, \& {Kraus}}]{Gallenne_2015_07_0}
{Gallenne}, A., {M{\'e}rand}, A., {Kervella}, P., {et~al.} 2015, \aap, 579, A68

\bibitem[{{Groenewegen}(2013)}]{Groenewegen_2013_02_0}
{Groenewegen}, M.~A.~T. 2013, \aap, 550, A70

\bibitem[{{Heintz}(1978)}]{Heintz_1978__0}
{Heintz}, W.~D. 1978, Geophysics and Astrophysics Monographs, 15

\bibitem[{{Henden}(1980)}]{Henden_1980_08_0}
{Henden}, A.~A. 1980, \mnras, 192, 621

\bibitem[{{Kervella} {et~al.}(2006){Kervella}, {M{\'e}rand}, {Perrin}, \&
  {Coud{\'e} Du Foresto}}]{Kervella_2006_03_0}
{Kervella}, P., {M{\'e}rand}, A., {Perrin}, G., \& {Coud{\'e} Du Foresto}, V.
  2006, \aap, 448, 623

\bibitem[{{Kervella} {et~al.}(2014){Kervella}, {Bond}, {Cracraft}, {Szabados},
  {Breitfelder}, {M{\'e}rand}, {Sparks}, {Gallenne}, {Bersier}, {Fouqu{\'e}},
  \& {Anderson}}]{Kervella_2014_12_0}
{Kervella}, P., {Bond}, H.~E., {Cracraft}, M., {et~al.} 2014, \aap, 572, A7

\bibitem[{{Kiss}(1998)}]{Kiss_1998_07_0}
{Kiss}, L.~L. 1998, \mnras, 297, 825

\bibitem[{{Kovtyukh} {et~al.}(2008){Kovtyukh}, {Soubiran}, {Luck}, {Turner},
  {Belik}, {Andrievsky}, \& {Chekhonadskikh}}]{Kovtyukh_2008_09_0}
{Kovtyukh}, V.~V., {Soubiran}, C., {Luck}, R.~E., {et~al.} 2008, \mnras, 389,
  1336

\bibitem[{{Leavitt} \& {Pickering}(1912)}]{Leavitt_1912_03_0}
{Leavitt}, H.~S., \& {Pickering}, E.~C. 1912, Harvard College Observatory
  Circular, 173, 1

\bibitem[{{Lindegren} {et~al.}(2018){Lindegren}, {Hern{\'a}ndez}, {Bombrun},
  {Klioner}, {Bastian}, {Ramos-Lerate}, {de Torres}, {Steidelm{\"u}ller},
  {Stephenson}, {Hobbs}, {Lammers}, {Biermann}, {Geyer}, {Hilger}, {Michalik},
  {Stampa}, {McMillan}, {Casta{\~n}eda}, {Clotet}, {Comoretto}, {Davidson},
  {Fabricius}, {Gracia}, {Hambly}, {Hutton}, {Mora}, {Portell}, {van Leeuwen},
  {Abbas}, {Abreu}, {Altmann}, {Andrei}, {Anglada}, {Balaguer-N{\'u}{\~n}ez},
  {Barache}, {Becciani}, {Bertone}, {Bianchi}, {Bouquillon}, {Bourda},
  {Br{\"u}semeister}, {Bucciarelli}, {Busonero}, {Buzzi}, {Cancelliere},
  {Carlucci}, {Charlot}, {Cheek}, {Crosta}, {Crowley}, {de Bruijne}, {de
  Felice}, {Drimmel}, {Esquej}, {Fienga}, {Fraile}, {Gai}, {Garralda},
  {Gonz{\'a}lez-Vidal}, {Guerra}, {Hauser}, {Hofmann}, {Holl}, {Jordan},
  {Lattanzi}, {Lenhardt}, {Liao}, {Licata}, {Lister}, {L{\"o}ffler},
  {Marchant}, {Martin-Fleitas}, {Messineo}, {Mignard}, {Morbidelli}, {Poggio},
  {Riva}, {Rowell}, {Salguero}, {Sarasso}, {Sciacca}, {Siddiqui}, {Smart},
  {Spagna}, {Steele}, {Taris}, {Torra}, {van Elteren}, {van Reeven}, \&
  {Vecchiato}}]{Lindegren_2018_08_0}
{Lindegren}, L., {Hern{\'a}ndez}, J., {Bombrun}, A., {et~al.} 2018, \aap, 616,
  A2

\bibitem[{{Luck} {et~al.}(2008){Luck}, {Andrievsky}, {Fokin}, \&
  {Kovtyukh}}]{Luck_2008_07_0}
{Luck}, R.~E., {Andrievsky}, S.~M., {Fokin}, A., \& {Kovtyukh}, V.~V. 2008,
  \aj, 136, 98

\bibitem[{{Madore} \& {Freedman}(2012)}]{Madore_2012_01_0}
{Madore}, B.~F., \& {Freedman}, W.~L. 2012, \apj, 744, 132

\bibitem[{{M{\'e}rand} {et~al.}(2006){M{\'e}rand}, {Kervella}, {Coud{\'e} Du
  Foresto}, {Perrin}, {Ridgway}, {Aufdenberg}, {Ten Brummelaar}, {McAlister},
  {Sturmann}, {Sturmann}, {Turner}, \& {Berger}}]{Merand_2006_07_0}
{M{\'e}rand}, A., {Kervella}, P., {Coud{\'e} Du Foresto}, V., {et~al.} 2006,
  \aap, 453, 155

\bibitem[{{Monnier} {et~al.}(2004){Monnier}, {Berger}, {Millan-Gabet}, \& {ten
  Brummelaar}}]{Monnier_2004_10_0}
{Monnier}, J.~D., {Berger}, J.-P., {Millan-Gabet}, R., \& {ten Brummelaar},
  T.~A. 2004, in SPIE Conference Series, ed. {W.~A.~Traub}, Vol. 5491, 1370

\bibitem[{{Monnier} {et~al.}(2007){Monnier}, {Zhao}, {Pedretti}, {Thureau},
  {Ireland}, {Muirhead}, {Berger}, {Millan-Gabet}, {Van Belle}, {ten
  Brummelaar}, {McAlister}, {Ridgway}, {Turner}, {Sturmann}, {Sturmann}, \&
  {Berger}}]{Monnier_2007_07_0}
{Monnier}, J.~D., {Zhao}, M., {Pedretti}, E., {et~al.} 2007, Science, 317, 342

\bibitem[{{Pecaut} \& {Mamajek}(2013)}]{Pecaut_2013_09_0}
{Pecaut}, M.~J., \& {Mamajek}, E.~E. 2013, \apjs, 208, 9

\bibitem[{{Pilecki} {et~al.}(2018){Pilecki}, {Gieren}, {Pietrzy{\'n}ski},
  {Thompson}, {Smolec}, {Graczyk}, {Taormina}, {Udalski}, {Storm}, {Nardetto},
  {Gallenne}, {Kervella}, {Soszy{\'n}ski}, {G{\'o}rski}, {Wielg{\'o}rski},
  {Suchomska}, {Karczmarek}, \& {Zgirski}}]{Pilecki_2018_07_0}
{Pilecki}, B., {Gieren}, W., {Pietrzy{\'n}ski}, G., {et~al.} 2018, \apj, 862,
  43

\bibitem[{{Planck Collaboration} {et~al.}(2018){Planck Collaboration},
  {Akrami}, {Arroja}, {Ashdown}, {Aumont}, {Baccigalupi}, {Ballardini},
  {Banday}, {Barreiro}, {Bartolo}, {Basak}, {Battye}, {Benabed}, {Bernard},
  {Bersanelli}, {Bielewicz}, {Bock}, {Bond}, {Borrill}, {Bouchet}, {Boulanger},
  {Bucher}, {Burigana}, {Butler}, {Calabrese}, {Cardoso}, {Carron},
  {Casaponsa}, {Challinor}, {Chiang}, {Colombo}, {Combet}, {Contreras},
  {Crill}, {Cuttaia}, {de Bernardis}, {de Zotti}, {Delabrouille}, {Delouis},
  {D{\'e}sert}, {Di Valentino}, {Dickinson}, {Diego}, {Donzelli}, {Dor{\'e}},
  {Douspis}, {Ducout}, {Dupac}, {Efstathiou}, {Elsner}, {En{\ss}lin},
  {Eriksen}, {Falgarone}, {Fantaye}, {Fergusson}, {Fernandez-Cobos}, {Finelli},
  {Forastieri}, {Frailis}, {Franceschi}, {Frolov}, {Galeotta}, {Galli},
  {Ganga}, {G{\'e}nova-Santos}, {Gerbino}, {Ghosh}, {Gonz{\'a}lez-Nuevo},
  {G{\'o}rski}, {Gratton}, {Gruppuso}, {Gudmundsson}, {Hamann}, {Handley},
  {Hansen}, {Helou}, {Herranz}, {Hivon}, {Huang}, {Jaffe}, {Jones}, {Karakci},
  {Keih{\"a}nen}, {Keskitalo}, {Kiiveri}, {Kim}, {Kisner}, {Knox},
  {Krachmalnicoff}, {Kunz}, {Kurki-Suonio}, {Lagache}, {Lamarre}, {Langer},
  {Lasenby}, {Lattanzi}, {Lawrence}, {Le Jeune}, {Leahy}, {Lesgourgues},
  {Levrier}, {Lewis}, {Liguori}, {Lilje}, {Lilley}, {Lindholm},
  {L{\'o}pez-Caniego}, {Lubin}, {Ma}, {Mac{\'{\i}}as-P{\'e}rez}, {Maggio},
  {Maino}, {Mandolesi}, {Mangilli}, {Marcos-Caballero}, {Maris}, {Martin},
  {Mart{\'{\i}}nez-Gonz{\'a}lez}, {Matarrese}, {Mauri}, {McEwen}, {Meerburg},
  {Meinhold}, {Melchiorri}, {Mennella}, {Migliaccio}, {Millea}, {Mitra},
  {Miville-Desch{\^e}nes}, {Molinari}, {Moneti}, {Montier}, {Morgante}, {Moss},
  {Mottet}, {M{\"u}nchmeyer}, {Natoli}, {N{\o}rgaard-Nielsen}, {Oxborrow},
  {Pagano}, {Paoletti}, {Partridge}, {Patanchon}, {Pearson}, {Peel}, {Peiris},
  {Perrotta}, {Pettorino}, {Piacentini}, {Polastri}, {Polenta}, {Puget},
  {Rachen}, {Reinecke}, {Remazeilles}, {Renzi}, {Rocha}, {Rosset}, {Roudier},
  {Rubi{\~n}o-Mart{\'{\i}}n}, {Ruiz-Granados}, {Salvati}, {Sandri},
  {Savelainen}, {Scott}, {Shellard}, {Shiraishi}, {Sirignano}, {Sirri},
  {Spencer}, {Sunyaev}, {Suur-Uski}, {Tauber}, {Tavagnacco}, {Tenti},
  {Terenzi}, {Toffolatti}, {Tomasi}, {Trombetti}, {Valiviita}, {Van Tent},
  {Vibert}, {Vielva}, {Villa}, {Vittorio}, {Wandelt}, {Wehus}, {White},
  {White}, {Zacchei}, \& {Zonca}}]{Planck-Collaboration_2018_07_0}
{Planck Collaboration}, {Akrami}, Y., {Arroja}, F., {et~al.} 2018, \aap, in
  press [e-prints: 1807.06205]

\bibitem[{{Proffitt} {et~al.}(2017){Proffitt}, {Evans}, {Winston}, {Gallenne},
  \& {Kervella}}]{Proffitt_2017_09_0}
{Proffitt}, C.~R., {Evans}, N.~R., {Winston}, E.~M., {Gallenne}, A., \&
  {Kervella}, P. 2017, in European Physical Journal Web of Conferences, Vol.
  152, 04003

\bibitem[{{Raskin} {et~al.}(2011){Raskin}, {van Winckel}, {Hensberge},
  {Jorissen}, {Lehmann}, {Waelkens}, {Avila}, {de Cuyper}, {Degroote},
  {Dubosson}, {Dumortier}, {Fr{\'e}mat}, {Laux}, {Michaud}, {Morren}, {Perez
  Padilla}, {Pessemier}, {Prins}, {Smolders}, {van Eck}, \&
  {Winkler}}]{Raskin_2011_02_0}
{Raskin}, G., {van Winckel}, H., {Hensberge}, H., {et~al.} 2011, \aap, 526, A69

\bibitem[{{Riess} {et~al.}(2016){Riess}, {Macri}, {Hoffmann}, {Scolnic},
  {Casertano}, {Filippenko}, {Tucker}, {Reid}, {Jones}, {Silverman},
  {Chornock}, {Challis}, {Yuan}, {Brown}, \& {Foley}}]{Riess_2016_07_0}
{Riess}, A.~G., {Macri}, L.~M., {Hoffmann}, S.~L., {et~al.} 2016, \apj, 826, 56

\bibitem[{{Riess} {et~al.}(2018){Riess}, {Casertano}, {Yuan}, {Macri},
  {Anderson}, {MacKenty}, {Bowers}, {Clubb}, {Filippenko}, {Jones}, \&
  {Tucker}}]{Riess_2018_03_0}
{Riess}, A.~G., {Casertano}, S., {Yuan}, W., {et~al.} 2018, \apj, 855, 136

\bibitem[{{Sana} \& {Evans}(2011)}]{Sana_2011_07_0}
{Sana}, H., \& {Evans}, C.~J. 2011, in IAU Symposium, Vol. 272, Active OB
  Stars: Structure, Evolution, Mass Loss, and Critical Limits, ed. C.~{Neiner},
  G.~{Wade}, G.~{Meynet}, \& G.~{Peters}, 474--485

\bibitem[{{Sandage} {et~al.}(2004){Sandage}, {Tammann}, \&
  {Reindl}}]{Sandage_2004_09_0}
{Sandage}, A., {Tammann}, G.~A., \& {Reindl}, B. 2004, \aap, 424, 43

\bibitem[{{Storm} {et~al.}(2011){Storm}, {Gieren}, {Fouqu{\'e}}, {Barnes},
  {Pietrzy{\'n}ski}, {Nardetto}, {Weber}, {Granzer}, \&
  {Strassmeier}}]{Storm_2011_10_0}
{Storm}, J., {Gieren}, W., {Fouqu{\'e}}, P., {et~al.} 2011, \aap, 534, A94

\bibitem[{{Szabados}(1977)}]{Szabados_1977_01_0}
{Szabados}, L. 1977, Commun. of the Konkoly Observatory Hungary, 70, 1

\bibitem[{{Szabados}(2003)}]{Szabados_2003__0}
{Szabados}, L. 2003, in Astronomical Society of the Pacific Conference Series,
  Vol. 298, GAIA Spectroscopy: Science and Technology, ed. {U.~Munari}, 237

\bibitem[{{Szabados} \& {Klagyivik}(2012)}]{Szabados_2012_09_0}
{Szabados}, L., \& {Klagyivik}, P. 2012, \apss, 341, 99

\bibitem[{{Taormina} {et~al.}(2018){Taormina}, {Pilecki}, \&
  {Smolec}}]{Taormina_2018_03_0}
{Taormina}, M., {Pilecki}, B., \& {Smolec}, R. 2018, in RR Lyrae 2017
  Conference proceedings [e-prints: 1803.10911]

\bibitem[{{ten Brummelaar} {et~al.}(2005){ten Brummelaar}, {McAlister},
  {Ridgway}, {Bagnuolo}, {Turner}, {Sturmann}, {Sturmann}, {Berger}, {Ogden},
  {Cadman}, {Hartkopf}, {Hopper}, \& {Shure}}]{ten-Brummelaar_2005_07_0}
{ten Brummelaar}, T.~A., {McAlister}, H.~A., {Ridgway}, S.~T., {et~al.} 2005,
  \apj, 628, 453

\bibitem[{{Wright} \& {Howard}(2009)}]{Wright_2009_05_0}
{Wright}, J.~T., \& {Howard}, A.~W. 2009, \apjs, 182, 205

\end{thebibliography}

%
%



\end{document}